\documentclass[11pt]{article}
\usepackage{Blois,epsfig}

\usepackage{amssymb}
\usepackage{graphicx}
\usepackage{bm}
\usepackage{citesort}
\usepackage{makeidx}
\usepackage{multicol}
\usepackage{bbm}
\usepackage{amsmath}
\usepackage{epsfig}
\long\def\comment#1{ }

\newcommand{\beq}{\begin{eqnarray}}
\newcommand{\eeq}{\end{eqnarray}}
\newcommand{\be}{\begin{eqnarray}}
\newcommand{\ee}{\end{eqnarray}}
\newcommand{\E}{Eq.~\eqref}

\newcommand{\abar}{\bar{\alpha}_s}

\newcommand{\mean}[1]{\left\langle #1 \right\rangle_\tau}

\newcommand{\dif}{{\rm d}}

\def\simge{\mathrel{%
   \rlap{\raise 0.511ex \hbox{$>$}}{\lower 0.511ex \hbox{$\sim$}}}}
\def\simle{\mathrel{
   \rlap{\raise 0.511ex \hbox{$<$}}{\lower 0.511ex \hbox{$\sim$}}}}
\def\bigs{\mathrel{
   \rlap{\raise 0.531ex \hbox{$>$}}{\lower 0.531ex \hbox{$<$}}}}

\def\empile#1\over#2{\mathrel{\mathop{\kern 0pt#1}\limits_{#2}}}

\def\del{\partial}                              

\newcommand{\x}{\bm x}
\newcommand{\y}{\bm y}

\newcommand{\z}{\bm z}

\begin{document}

\vspace*{4cm}
\title{POMERON LOOPS \& DUALITY IN HIGH ENERGY QCD
 }

\author{Edmond IANCU
}

\address{Service de Physique Th\'eorique, CEA/DSM/SPhT,
CE Saclay, F-91191 Gif-sur-Yvette, France}

\maketitle\abstracts{I discuss the physical picture underlying the
evolution equations with Pomeron loops recently derived in
multicolor QCD at high energy and qualitatively explain the notion
of `self--duality'.}

\begin{figure}[t]
  \begin{center}
    \centerline{\epsfxsize=3.6cm\epsfbox{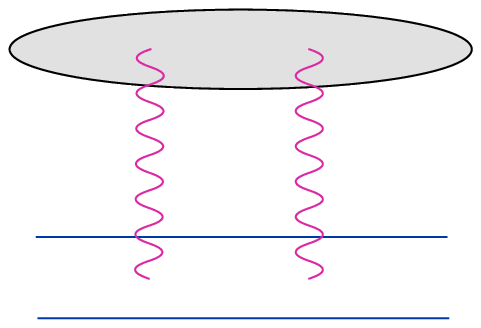}
    \hspace{0.3cm}\epsfxsize=3.6cm\epsfbox{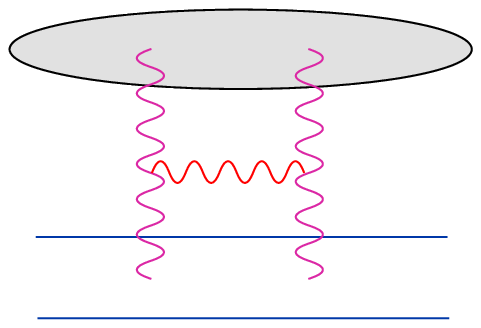}
    \hspace{0.3cm}\epsfxsize=3.9cm\epsfbox{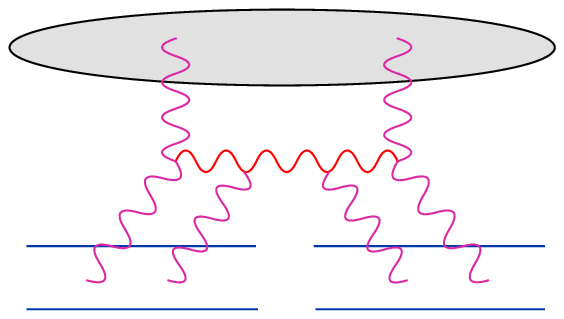}
    \hspace{0.3cm}\epsfxsize=3.6cm\epsfbox{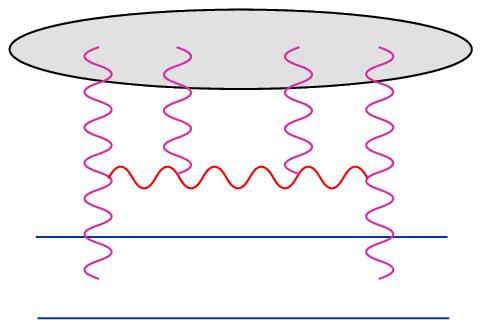}}
     \vspace*{0.1cm}
{\small\sl \hspace*{-.4cm} (a)\hspace{3.5cm} (b)\hspace{3.8cm}(c)
 \hspace{3.4cm} (d)}
    \caption{\sl Elementary processes which occur
    in one--step evolution in QCD
    at high energy: (a) tree--level process; (b) standard BFKL
    evolution; (c) $2\to 4$ gluon splitting; (d) $4\to 2$ gluon
    merging.
    \label{ONE_STEP}}
   \end{center}
\end{figure}

The rapid rise of the gluon density in a hadronic wavefunction
with increasing energy together with the phenomenon of gluon
saturation open the way towards a perturbative treatment of
high--energy scattering in QCD in the vicinity of the unitarity
limit \cite{CGC}. To that aim, one needs evolution equations which
include all the microscopic processes contributing in perturbative
QCD to the growth of the gluon distribution and its eventual
saturation. These processes have been discussed in a different
article in these Proceedings \cite{Blois1} and are also summarized
in Fig. \ref{ONE_STEP}.  They include the {\em BFKL evolution}
(Fig. \ref{ONE_STEP}.b), which involves a $2\to 2$ gluon vertex
and leads to the rapid growth of the gluon density with increasing
energy, together with {\em gluon splitting} vertices $2\to n$ with
$n>2$ (Fig. \ref{ONE_STEP}.c) and {\em gluon recombination}
vertices $n\to 2$ (Fig. \ref{ONE_STEP}.d). One must emphasize here
that {\em all} such processes
--- in fact, all the gluon number changing vertices $m\to n$ with
arbitrary $m,\,n\ge 2$ --- are important for computing scattering
amplitudes at very high energy \cite{IT04}, and this even when the
projectile is a relatively simple object, like a {\em  color
dipole} (the interesting case for deep inelastic scattering at
small $x$). Indeed, the non--linear effects responsible for gluon
saturation and unitarization involve higher $n$--point correlation
functions (cf. Fig. \ref{ONE_STEP}.d), which in turn are generated
through splitting processes in the low density regime (cf. Fig.
\ref{ONE_STEP}.c). In Fig. \ref{ONE_STEP}, we have also indicated
some simple scattering processes involving one or two external
dipoles that we shall use below to probe the gluon evolution in
the target.

Gluon--number changing vertices started to be computed in
perturbative QCD in the early nineties \cite{ewerz}, but their
systematic inclusion into evolution equations turned out to be a
complicated problem, whose general solution is not yet known. The
{\em merging} vertices are presently included in the functional
JIMWLK equation \cite{CGC}, whereas the {\em splitting} vertices
are rather described by the {\em dual} version  of the JIMWLK
Hamiltonian \cite{KL05} (see below for the notion of `duality').
But the complete evolution Hamiltonian, which should include both
splitting and mergings, and therefore be {\em self--dual}
\cite{KL05,BIIT05}, is not yet fully known (see however Refs.
\cite{BREM}).

The desiderata of including both splitting and merging has been
nevertheless achieved \cite{IT04,MSW05} within the multicolor
limit $N_c\to \infty$, which allows for important technical
simplifications and also for the use of a suggestive dipole
language \cite{AM94,IM031} (see also Refs.
\cite{LL05,Braun05,BIIT05}).

\begin{figure}[t]
    \centerline{\epsfxsize=4.cm\epsfbox{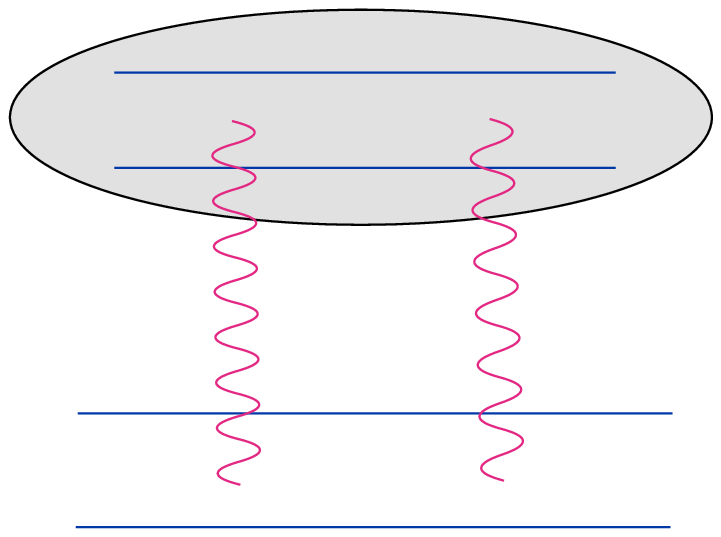}
    \hspace{0.7cm}\epsfxsize=4.cm\epsfbox{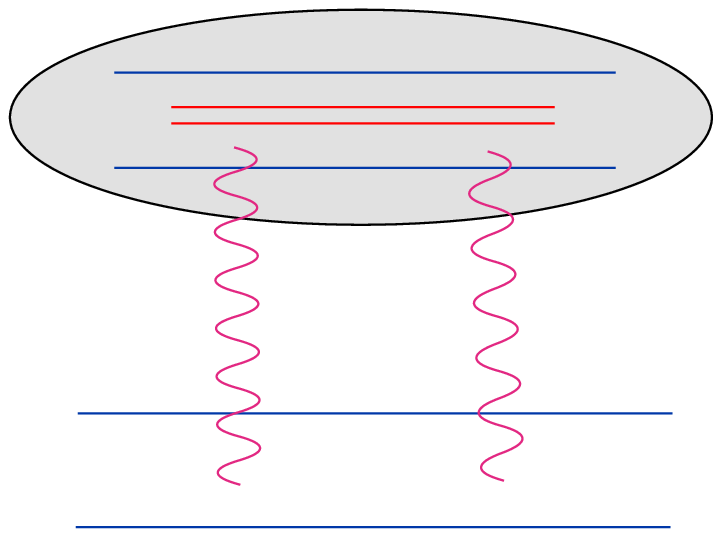}
    \hspace{0.7cm}\epsfxsize=4.cm\epsfbox{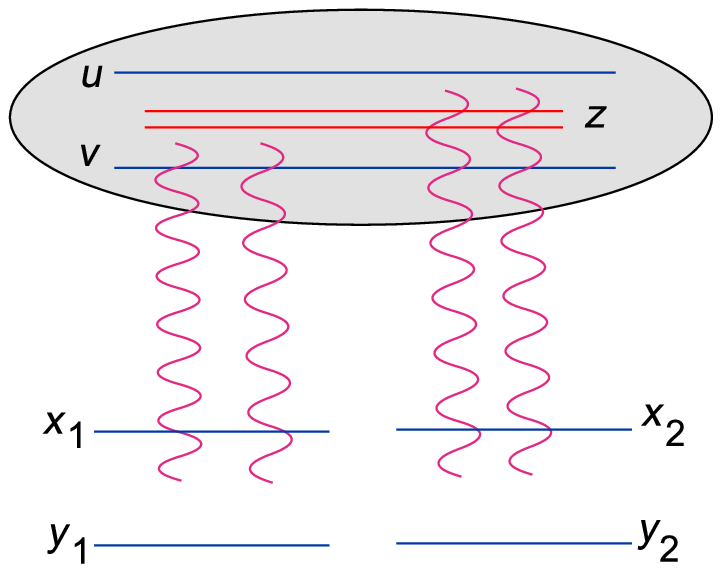}}
      \vspace*{0.1cm}
{\small\sl \hspace{2.7cm} (a)\hspace{4.3cm} (b)\hspace{4.3cm}(c) }
    \caption{\sl Target evolution in the dipole picture. \label{TARGET}}
\end{figure}

Namely, at large $N_c$ and in the dilute regime, the gluons in the
target wavefunction can be effectively replaced by color dipoles,
which interact with the external dipoles from the projectile via
two gluon exchanges (compare in this respect the processes in
Figs. \ref{ONE_STEP}.a and \ref{TARGET}.a). Then the BFKL
evolution together with the $2\to 4$ gluon splitting\footnote{At
large $N_c$, and for a dipole projectile, one can ignore all the
splitting vertices except for the vertex $2\to 4$.} (as previously
depicted in Figs. \ref{ONE_STEP}.b and c) can be represented as
one target dipole splitting into two, followed by the interaction
between the child dipoles and the projectile (cf. Figs.
\ref{TARGET}.b and c). Note that, whereas the scattering with a
single external dipole serves as a measure of the {\em average
dipole number density} in the target, and thus of the standard
BFKL evolution (cf. Fig. \ref{TARGET}.b), on the other hand, the
scattering with two external dipoles (cf. Fig. \ref{TARGET}.c) can
measure both dipoles produced after a splitting, and thus probes
the {\em dipole number fluctuations} in the target.

\begin{figure}[t]
    \centerline{
    \epsfxsize=4.cm\epsfbox{Fig2a.eps}
    \hspace{0.7cm}\epsfxsize=4.cm\epsfbox{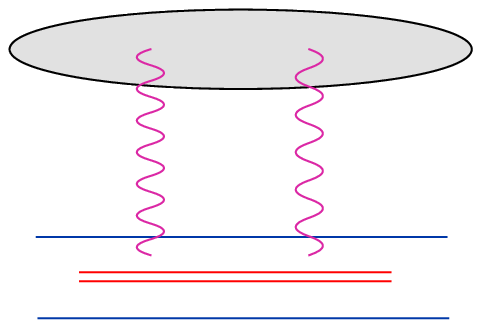}
    \hspace{0.7cm}\epsfxsize=4.cm\epsfbox{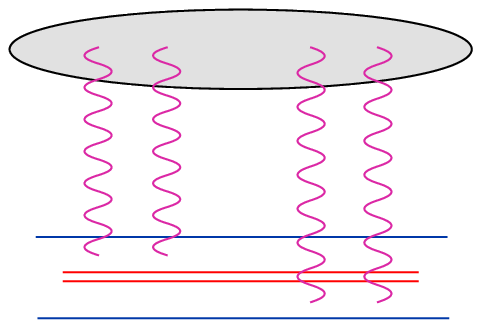}}
      \vspace*{0.1cm}
{\small\sl \hspace{2.7cm} (a)\hspace{4.3cm} (b)\hspace{4.3cm}(c) }
     \caption{\sl Projectile evolution in the dipole picture.
    \label{PROJ}}
\end{figure}

But the dipole picture breaks down at high density \cite{AM94},
and thus cannot be used  \cite{IST05} to also describe the $4\to
2$ recombination process in Fig. \ref{ONE_STEP}.d. However, by
viewing the diagram  in Fig. \ref{ONE_STEP}.d upside down, it
becomes clear that a merging process in the {\em target} can be
alternatively interpreted as a splitting process in the {\em
projectile} ; since the latter is dilute, the dipole picture can
then be used to describe its evolution. The two diagrams in Figs.
\ref{PROJ}.b and c represent the same processes as previously
shown in Figs. \ref{ONE_STEP}.b and d, respectively, but now from
the perspective of projectile evolution. In particular, the
diagram in Fig. \ref{PROJ}.c describes the simultaneous scattering
of both child dipoles in the projectile, which is the
unitarization mechanism at high energy. We see that the {\em
unitarity corrections} correspond to either {\em gluon
saturation}, cf. Fig. \ref{ONE_STEP}.d, or to {\em multiple
scattering}, cf. Fig. \ref{PROJ}.c, according to the frame in
which we choose to view the evolution.

\comment{Note also that, from the viewpoint of projectile
evolution, the process in Fig. \ref{ONE_STEP}.c (which
corresponds, as we have seen, the splitting in the target) appears
as a recombination effect, i.e., an interaction between two
dipoles in the projectile wavefunction before the scattering with
the target. However, as already mentioned, this interaction does
not lead to the actual fusion of two dipoles into one.}

The previous considerations show that, in order to construct
evolution equations which include both splitting and merging at
large $N_c$, it is in fact sufficient to know the elementary {\em
dipole splitting vertex}, that is, the probability density for an
original color dipole to split into two new dipoles per unit
rapidity. This has been first computed by Al Mueller \cite{AM94},
and reads
   \be \label{dPsplit} \frac{dP}{d\tau}\,=\,
 \frac{\alpha_s N_c}{2\pi^2}\,{\mathcal M}({\bm{x}},{\bm{y}},{\bm z})\,
 {d^2\z}\,,\qquad {\mathcal M}({\bm{x}},{\bm{y}},{\bm z})\,\equiv\,
 \frac{(\x-\y)^2}{(\x-\z)^2(\y-\z)^2 }\,
  ,\ee
where $d\tau$ is the rapidity increment ($\tau\sim\ln s$ is the
rapidity gap between the projectile and the target), the pair
$({\bm{x}},{\bm{y}})$ denotes the transverse coordinates of the
quark and the antiquark in the original dipole, and
$({\bm{x}},{\bm{z}})$ and $({\bm{z}},{\bm{y}})$ refer similarly to
the two final dipoles. The latter have a common leg at $\z$, which
physically represents the transverse coordinate of the emitted
gluon.

By using Eq.~(\ref{dPsplit}) together with the picture of {\em
projectile evolution}, cf. Fig. \ref{PROJ}, one can immediately
deduce the following evolution equation for the average scattering
amplitude $\langle T(\x,\y) \rangle_\tau$ for a projectile made
with a single dipole $({\bm{x}},{\bm{y}})$ (with $\bar{\alpha}_s =
{\alpha}_s N_c/\pi$) \cite{BK} :
  \be
 \label{B1}\frac{\del}{\del \tau}\langle
 T(\x,\y)
 \rangle_\tau =\frac{\bar{\alpha}_s}{2\pi} \!\int_{\z}\,
{\mathcal M}({\bm{x}},{\bm{y}},{\bm z})\,
 \Big \langle - T(\x,\y) +  T(\x,\z) + T(\z,\y)
  - \langle T(\x,\z) T(\z,\y)
  \Big\rangle_\tau.
  \ee
Together, the first three terms in the r.h.s., which are linear in
$\langle T \rangle$, represent the BFKL evolution, that is, the
scattering of only one of the two child dipoles (cf. Fig.
\ref{PROJ}.b) and the reduction in the probability that the
incoming dipole remains in its original state (the `virtual term'
$- \langle T(\x,\y)\rangle$). The last, non--linear, term $\langle
T(\x,\z) T(\z,\y) \rangle_\tau$ describes the simultaneous
scattering of both child dipoles, cf. Fig. \ref{PROJ}.c. So long
as the scattering is weak, $\langle T \rangle_\tau\ll 1$, this
last term is negligible and the amplitude rises very fast
(exponentially in $\tau$), according to the BFKL equation. But
when $\langle T \rangle_\tau\sim 1$, the non--linear term becomes
important and tames the growth. Thus, as anticipated, the multiple
scattering is the mechanism leading to unitarization.

But in order to study this mechanism in detail, one also needs the
equation satisfied by the 2--dipole amplitude  $\langle T(\x,\z)
T(\z,\y) \rangle_\tau$, or, more generally, $\langle T
(\bm{x}_1,\bm{y}_1) T(\bm{x}_2,\bm{y}_2) \rangle_\tau$. This
involves, first, an evolution similar to Eq.~(\ref{B1}) for any of
the two incoming dipoles $(\bm{x}_1,\bm{y}_1)$ or
$(\bm{x}_2,\bm{y}_2)$ (while the other one is simply a spectator).
This will contribute terms linear in $\langle T T\rangle_\tau$
(the BFKL terms) together with unitarity corrections proportional
to $\langle T T T\rangle_\tau$. Besides, the two external dipoles
can `see' a fluctuation in the gluon in the target, that is, they
can directly probe the $2\to 4$ splitting vertex in Fig.
\ref{ONE_STEP}.c. To estimate this additional effect, it is
preferable to adopt the point of view of {\em target evolution},
cf. Fig. \ref{TARGET}.c, which implies \cite{IT04}
 \begin{align}\label{T2evol}
    \frac{\partial \mean{T(\bm{x}_1,\bm{y}_1) T(\bm{x}_2,\bm{y}_2)}}
    {\partial \tau}
    \bigg|_{\rm fluct}\!\!\! =
    \left(\frac{\alpha_s}{2\pi}\right)^2
    \frac{\abar}{2 \pi}\!
    \int\limits_{\bm{u},\bm{v},\bm{z}}&
    \mathcal{M}(\bm{u},\bm{v},\bm{z})\,
    \mathcal{A}_0(\bm{x}_1,\bm{y}_1|\bm{u},\bm{z})\,
    \mathcal{A}_0(\bm{x}_2,\bm{y}_2|\bm{z},\bm{v})\,
    \nonumber \\
    &\times\nabla_{\bm{u}}^2 \nabla_{\bm{v}}^2\, \mean{T(\bm{u},\bm{v})}.
\end{align}
The r.h.s. of this equation should be read as follows: A dipole
$(\bm{u},\bm{v})$ from the target splits into two new dipoles
$(\bm{u},\bm{z})$ and $(\bm{z},\bm{v})$ with probability
$(\abar/2\pi) \mathcal{M}(\bm{u},\bm{v},\bm{z})$ (cf.
\E{dPsplit}), then the two child dipoles scatter off the external
dipoles, with an
amplitude\footnote{$\alpha_s^2{\cal{A}}_0(\bm{x},\bm{y}|\bm{u},\bm{v})$
is the standard amplitude for dipole--dipole scattering via
two--gluon exchange \cite{IT04}.} $\alpha_s^2\mathcal{A}_0$ for
each scattering. Finally, the `amputated' amplitude
$(1/\alpha_s^2)\nabla_{\bm{u}}^2 \nabla_{\bm{v}}^2\,
\mean{T(\bm{u},\bm{v})}$ is, up to a normalization factor, the
average dipole number density $\mean{n(\bm{u},\bm{v})}$ in the
target.

Note that, although suppressed by an explicit factor $\alpha_s^2$,
the contribution of the dipole number fluctuations to the
evolution of $\langle T T\rangle_\tau$ is in fact a {\em dominant}
effect in the {\em very dilute} regime where $\langle T
\rangle_\tau\simle \alpha_s^2\,$: Indeed, in this regime $\langle
T \rangle \langle T \rangle < \alpha_s^2 \langle T \rangle$, hence
the dominant contribution to the rise of $\langle T T\rangle_\tau$
comes from fluctuations, via Eq.~(\ref{T2evol}). This conclusion
is in agreement with the qualitative analysis of the fluctuations
in the companion article, Ref. \cite{Blois1}.

It is now clear that the evolution equations for dipole scattering
amplitudes form an hierarchy, in which the general equation reads
as follows (with $t=\abar \tau$ and $T^{(n)}\equiv \langle
T(1)T(2)\dots T(n)\rangle_\tau$)
 \be
 \label{PLOOP}
  \frac{\dif T^{(n)}}{d t}
  = T^{(n)} - T^{(n+1)} + \alpha_s^2 T^{(n-1)},
 \ee
in schematic notations which ignore combinatoric factors and the
non--locality of the various vertices in the transverse space. The
first two terms in the r.h.s. correspond to the BFKL evolution and
the unitarity corrections, respectively, while the last term
describes dipole number fluctuations in the target wavefunction.
The iterative solution to these equations can be given a
diagrammatic representation in terms of `BFKL Pomerons' (the
Green's function of the BFKL equation) which split and merge with
each other, and thus form {\em Pomeron loops} \cite{BIIT05}.

An hierarchy like that in Eq.~(\ref{PLOOP}) can be exactly
represented by a Langevin equation with a specific noise term. In
statistical physics, this equation is known as the {\em stochastic
FKPP equation} (sFKPP), and has the following generic structure
(see Ref. \cite{IT04} for details) :
 \be \frac{\dif T}{d t} = T - T^2+\sqrt{\alpha_s^2\, T}\,\nu(t)
    \qquad {\rm with} \qquad
    \langle \nu(t) \nu(t') \rangle = \delta(t-t').\ee
Such an equation represents a convenient starting point for
numerical simulations, and particular limits of it have been
already studied in this way \cite{GS05}.

We shall conclude this brief presentation with a qualitative
explanation of the concept of {\em self--duality} in the context
of the high--energy QCD evolution \cite{KL05,BIIT05,BREM}. The
seemingly trivial fact that a same $n\to 2$ vertex can be viewed
as either merging in the target or splitting in the projectile has
the physically important consequence that, in order to be {\em
boost invariant}, the evolution equations for the scattering
amplitudes must include the effects of {\em both} splitting and
merging. Indeed, physical quantities must come out the same
whatever is the system that we choose the evolve (the projectile
or the target), since such a choice is tantamount to using a
particular Lorentz frame. Let's then assume that the {\em
evolution Hamiltonian} $H$
--- the operator which governs the one--step evolution of the
gluon fields in a hadronic wavefunction --- contains a particular
{\em merging} vertex  $n\to 2$. When acting on the gluon fields in
the {\em target}, this particular term in $H$ generates an
evolution for the scattering amplitudes which is equivalent to
that that would be produced by a specific {\em splitting} vertex
(the mirror, or `dual', image of the original merging vertex)
acting on the gluons in the {\em projectile}. Hence, in order to
yield the same physical result when directly applied to the
projectile, the Hamiltonian must also include the specific
splitting vertex mentioned above, and therefore be {\em
self--dual}. The precise duality transformation relating merging
to splitting vertices depends upon the detailed structure of the
scattering operator, and can be found in
Refs.\,\cite{KL05,BIIT05,BREM}. The Hamiltonian underlying the
evolution equations with Pomeron loops presented above is indeed
self--dual \cite{BIIT05}, and so is also the more general
Hamiltonian in Refs.\,\cite{BREM}.

\end{document}